# A Universal Receiver for Uplink NOMA Systems

(*Invited paper*)

Xiangming Meng, Yiqun Wu, Chao Wang, and Yan Chen
Huawei Technologies, Co. Ltd.
Email: {mengxiangming1, wuyiqun, wangchao78, bigbird.chenyan}@huawei.com

*Abstract* — **Given its capability in efficient radio resource sharing, non-orthogonal multiple access (NOMA) has been identified as a promising technology in 5G to improve the system capacity, user connectivity, and scheduling latency. A dozen of uplink NOMA schemes have been proposed recently and this paper considers the design of a universal receiver suitable for all potential designs of NOMA schemes. Firstly, a general turbo-like iterative receiver structure is introduced, under which, a universal expectation propagation algorithm (EPA) detector with hybrid parallel interference cancellation (PIC) is proposed (EPA in short). Link-level simulations show that the proposed EPA receiver can achieve superior block error rate (BLER) performance with implementation friendly complexity and fast convergence, and is always better than the traditional codeword level MMSE-PIC receiver for various kinds of NOMA schemes.**

***Keywords—5G; NOMA; Iterative receiver; Universal EPA receiver***

## I. Introduction

The future fifth generation (5G) wireless networks are expected to support massive connectivity, low latency as well as better coverage [1]. Non-orthogonal multiple access (NOMA) technique has been identified as a promising radio resource sharing technology in 5G to help improving system capacity, user connectivity, and service latency. Generally, a NOMA transmitter maps a stream of coded binary bits of a user (UE) to the available transmission resources by some user-specific operations to facilitate decoding of the superposed multi-user data at the receiver side with reasonable complexity. For the design of the user-specific operations, both power domain and code domain user separation schemes have been proposed from academic and industry [1]-[6]. The NOMA concept together with 15 different schemes were proposed as candidates for 3GPP Rel-14 NR (New Radio) study in early 2016, such as sparse code multiple access (SCMA) [7][8], pattern division multiple access (PDMA) [9], interleave division multiple access (IDMA) [10], interleave grid multiple access (IGMA) [11], and multi-user shared multiple access (MUSA) [12]. In Rel-15, a dedicated NOMA study item was approved and kicked off since Feb 2018, which mainly focuses on the study of the uplink NOMA transceiver design and some related procedures. For the progress of the standardization of NOMA, the interested readers are referred to [3] for more details.

In this paper, we focus on the design of a universal receiver that is suitable for all uplink NOMA schemes. Towards this end, we firstly adopt a unified system model to describe the uplink NOMA system mathematically, based on which, we introduced a general turbo-like iterative receiver structure that supports hybrid (soft and hard) and parallel interference cancellation (PIC) between symbol domain multi-user (MU) detector and bit domain channel decoding. Among the potential candidates of MU detector, we propose the expectation propagation algorithm (EPA) [13][14] as a simplification of the MPA receiver and an enhancement for the elementary signal estimator (ESE) and minimum mean square error (MMSE) receivers. We then derive the detailed algorithmic implementation of the universal EPA receiver for all potential NOMA schemes based on a unified factor graph representation. The proposed EPA receiver is implemented in a chip-by-chip manner and the dominant complexity is of the order $\mathcal{O}(N_r^3 L)$, where $L$ is the spreading factor and $N_r$ is the number of receive antennas. Link level simulation results demonstrate its superiority in the block error rate (BLER) performance for various kinds of NOMA schemes (either spreading or non-spreading NOMA schemes), compared with the traditional codeword level MMSE-PIC receiver.

## II. Unified Receiver Design

### *A. System Model*

Consider an uplink NOMA system where there are $K$ independent single-antenna UEs and the base station employs $N_r$ receive antennas. Without loss of generality, we assume that each user transmits with one layer on a group of $L$ resource elements (RE). For spreading-based NOMA schemes, $L$ is referred to as spreading factor. For NOMA schemes without spreading, there is $L = 1$. The cases where each user's data takes more than one layer can be straight forward extension.

Specifically, for each UE $k, k = 1, \ldots, K$ , the binary information bits $\mathbf{b}_k$ are first encoded by a feedforward error correction (FEC) encoder with coding rate $R_k$. To randomize the inter-user/inter-cell interference, bit-level interleaver and/or scrambler can be applied in practice. Then, to generate NOMA signals whose modulation size is $M$, every $J = \log_2 M$ coded bits $\mathbf{c}_k = \left[c_{k,1}, c_{k,2}, \ldots, c_{k,J}\right]^T$ are mapped into an $L$-dimensional complex-valued symbol vector $\mathbf{x}_k = [x_{k1}, x_{k2}, \ldots, x_{kL}]^T$, where $(\cdot)^T$ denotes matrix or vector transpose. This step is also referred to as bits-to-symbols mapping. After passing the wireless channel, the received signal of the $r$-th antenna can be written as

$$\mathbf{y}_r = \sum_{k=1}^{K} \mathrm{diag}(\mathbf{h}_{k,r})\mathbf{x}_k + \mathbf{w}_r, \tag{1}$$

where $\mathbf{y}_r \in \mathbb{C}^{L\times 1}$ is the received symbol vector, $\mathbf{h}_{k,r} \in \mathbb{C}^{L\times 1}$ is the effective channel coefficient vector between user $k$ and $r$-th antenna, $\mathbf{x}_k \in \mathbb{C}^{L\times 1}$ is the transmitted symbol vector of user $k$, $\mathbf{w}_r$ is the associated additive complex Gaussian noise with power of $\sigma^2$, i.e., $\mathbf{w}_r \sim CN(\mathbf{w}_r; \mathbf{0}, \sigma^2\mathbf{I})$.

The observation equation in (1) is universal for all NOMA schemes, but the bits-to-symbols mapping step can be realized differently. For instance, if the symbol vector $\mathbf{x}_k$ is obtained by sequence spreading after traditional QAM modulation, then the effective channel $\mathbf{h}_{k,r}$ will take into account both the channel response of $r$-th receive antenna $\tilde{\mathbf{h}}_{k,r}$ and the user-specific spreading sequence $\mathbf{s}_k \in \mathbb{C}^{L\times 1}$, i.e., $\mathbf{h}_{k,r} = \tilde{\mathbf{h}}_{k,r} \odot \mathbf{s}_k$, where $\odot$ denotes component-wise multiplication, while $\mathbf{x}_k = [x_{k1}; \dots; x_{kL}] \in \mathbb{C}^{L\times 1}$ is a $L$ times repetition of the transmitted symbol $x_k$, i.e., $x_{kl} = x_k, l = 1, .., L$. On the other hand, if the symbol vector $\mathbf{x}_k$ is obtained by direct bits-to-symbol-vector mapping, such as SCMA [7], the effective channel $\mathbf{h}_{k,r}$ is the channel response of $r$-th receive antenna and $\mathbf{x}_k = [x_{k1}; \dots; x_{kL}] \in \mathbb{C}^{L\times 1}$ is the $L$-dimentional transmitted symbol vector selected from a predefined codebook. Note that if sparse pattern is applied, some elements of $\mathbf{x}_k$ are set to zero. The overall observations from $N_r$ receive antennas can be written as

$$\mathbf{y} = \mathbf{H}\mathbf{x} + \mathbf{w}, \tag{2}$$

where $\mathbf{y} = [\mathbf{y}_1; \dots; \mathbf{y}_{N_r}]$, $\mathbf{x} = [\mathbf{x}_1; \dots; \mathbf{x}_K]$, $\mathbf{w} = [\mathbf{w}_1; \dots; \mathbf{w}_{N_r}]$ and $\mathbf{H}$ is the overall effective channel matrix.

### B. Turbo-like Receiver Structure

A general high-level NOMA receiver structure has been agreed in the recent 3GPP meeting RAN1#92bis [15], which consists of symbol-level detector, FEC decoder, and iterative interference cancellation (IC) between the two. The IC can be hard or soft and can be implemented in serial (SIC) or parallel (PIC). Compared with SIC, PIC does not suffer from the error propagation problem and has low latency due to its parallelism. In addition, soft IC achieves better performance than hard IC due to the use of soft information. As a result, a turbo-like structure with hybrid PIC was proposed in [8] [16] as shown in Fig. 1, which takes advantages of both pure soft and pure hard IC schemes and achieves the best performance without much increase in complexity, which is evaluated in [16].

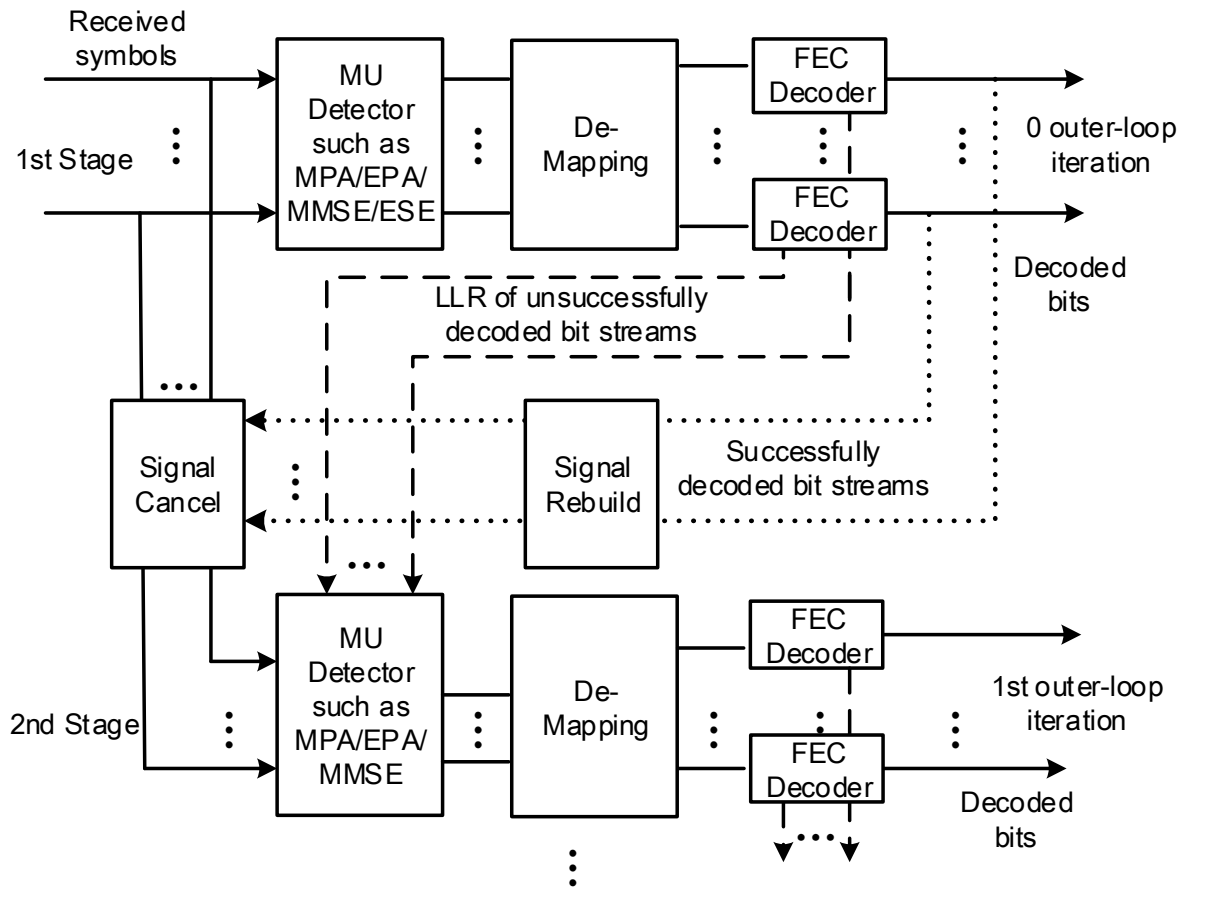

Fig. 1. Unified receiver structure with MU detector and hybrid PIC.

Specifically, as shown in Fig. 1, outer-loop iterations are adopted between the MU detector and the FEC decoder in a turbo manner with the hybrid soft and hard PIC, i.e., for users that are successfully decoded (passed the cyclic redundancy check (CRC)), their signals are reconstructed from the decoded information bits and canceled from the overall received signals; while for those users that are not successfully decoded, extrinsic log-likelihood ratios (LLRs) $\lambda^0(c_{k,j})$ are fed back as inputs to the MU detector. Note that in the unified receiver structure in Fig. 1, various kinds of MU detectors, e.g., MPA [18], ESE [10], or MMSE [17], can be applied and herein we consider the EPA detector as described in Section II.D. In particular, the number of outer-loop (OL) iterations $T_{\text{outer}}$ is defined as the number of information exchanges between the MU detector and the FEC decoder. Note that one time of OL iteration involves one time of parallel FEC decoding.

### C. Factor Graph Representation

Factor graph [18] is a kind of bipartite graph which represents the factorization of the joint probability distribution. As a natural graphical description, factor graph not only provides an intuitive representation, but also, perhaps, more importantly, provides a unified framework for deriving efficient inference algorithms. In fact, a wide variety of algorithms developed in the artificial intelligence, signal processing, and wireless communications communities may be derived as specific instances of the MPA operating in an appropriately chosen factor graph. For more details on factor graph and MPA, the interested readers are referred to [18].

The general factor graph representation for NOMA is shown in Fig. 2, which contains $K$ variable nodes (VNs) $\mathbf{x}_k$ and $L$ likelihood factor nodes (FNs, also known as function nodes) $f_l$, and $K$ prior FN $\Delta_k$. The VNs $\mathbf{x}_k$ represent the transmitted symbols and the prior FN $\Delta_k$ represents the prior distribution $P_0(\mathbf{x}_k)$ of $\mathbf{x}_k$, which can be computed via the feedback LLRs $\lambda^0(c_{k,j}), j = 1, \dots, J$ (if no feedback is available, $\lambda^0(c_{k,j}) = 0$) from the FEC decoder, i.e.,

$$P_0(\mathbf{x}_k) = \prod_{j=1}^{J} \frac{e^{c_{k,j}\lambda^0(c_{k,j})}}{1 + e^{c_{k,j}\lambda^0(c_{k,j})}}. \tag{3}$$

The likelihood FN $f_l$ represents the likelihood probability distribution $p(\bar{\mathbf{y}}_l|\mathbf{x})$, where $\bar{\mathbf{y}}_l = [y_l^1, \dots, y_l^{N_r}] \in \mathbb{C}^{N_r\times 1}$ is the observation vector at the $l$-th RE. Note that the FN $f_l$ has connection with $\mathbf{x}_k$ if and only if the $l$th element of $\mathbf{x}_k$, $x_{kl} \neq 0$. For ease of notation, let $V(k) = \{l: x_{kl} \neq 0\}$ denote the neighboring FNs of VN $\mathbf{x}_k$, and $F(l) = \{k: x_{kl} \neq 0\}$ denote the neighboring VNs of the FN $f_l$. The cardinalities of $V(k)$ and $F(l)$ are assumed to be $|V(k)| = d_v, k = 1, \dots, K, |F(l)| = d_f, l = 1, \dots, L$, respectively. As a result, the variables nodes connected with FN $f_l$ can be represented as a vector $\bar{\mathbf{x}}_l = [x_{kl}|k \in F(l)] \in \mathbb{C}^{d_f\times 1}$. The linear observation equation corresponding to FN $f_l$ is

$$\bar{\mathbf{y}}_l = \bar{\mathbf{H}}_l\bar{\mathbf{x}}_l + \bar{\mathbf{w}}_l \tag{4}$$

where $\bar{\mathbf{H}}_l \in \mathbb{C}^{N_r\times d_f}$ and $\bar{\mathbf{w}}_l \in \mathbb{C}^{N_r\times 1}$ are the corresponding channel matrix and additive Gaussian noise vector, respectively. Then, the likelihood distribution $p(\bar{\mathbf{y}}_l|\mathbf{x})$ can be written as

$$p(\bar{\mathbf{y}}_l|\mathbf{x}) \equiv P(\bar{\mathbf{y}}_l|\bar{\mathbf{x}}_l) = CN(\bar{\mathbf{y}}_l; \bar{\mathbf{H}}_l\bar{\mathbf{x}}_l, \sigma^2\mathbf{I}). \tag{5}$$

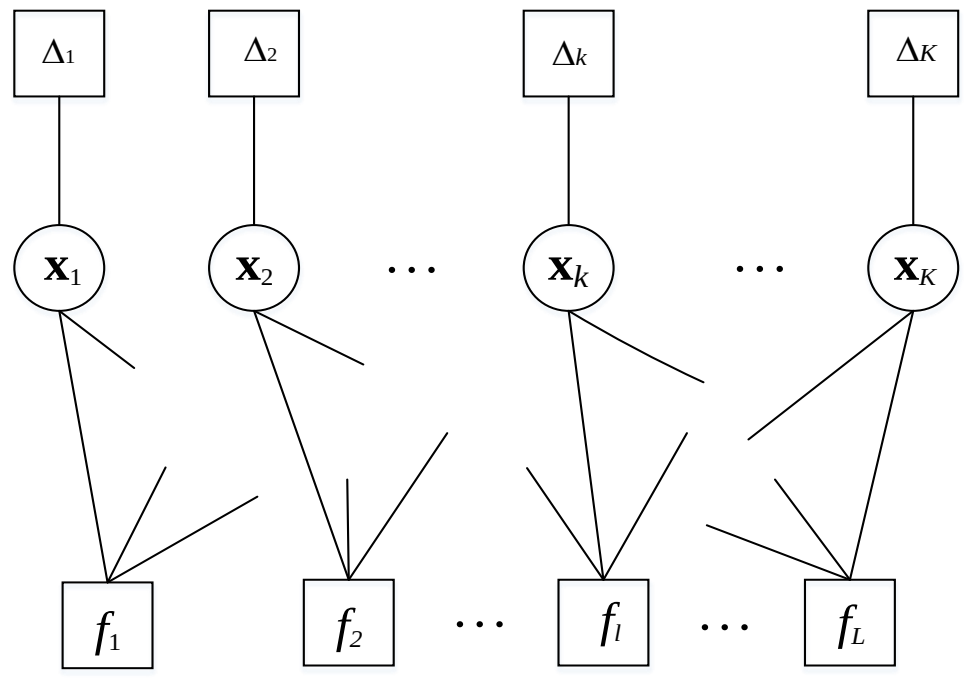

Fig. 2. Unified Factor Graph of NOMA.

Several comments worth mentioning for factor graph are as follows. First, since the adopted turbo-like receiver structure in Fig. 1 decouples the MU detector and FEC decoder, the factor graph in Fig. 2 does not include the FEC encoding function though it can be also included as a whole. Second, the specific factor graph differs for different kinds of NOMA schemes. For example, for non-spreading NOMA, i.e.,$L = 1$, it is a tree graph without cycles, while for non-sparse spreading, it is a fully-connected dense cyclic graph, and for sparse-spreading NOMA, it is a sparse cyclic graph which enables the application of near-maximum likelihood MPA detector [18]. Finally, there is much flexibility in the factor graph representations, which means that different factorizations of the joint distribution will lead to different representations.

### D. EPA Detector

Expectation propagation algorithm (EPA) is a well-known approximate Bayesian inference technique which approximates the target distribution $p$ with a simpler exponential family distribution $q$, where the Kullback-Leibler divergence $KL(p||q)$ is minimized [14]. It has already been widely used in the field of machine learning. Intuitively, EPA can be viewed as an operation to project the target distribution $p$ into the exponential family distribution set $\Phi$, i.e., $\mathrm{Proj}_{\Phi}(p) = \arg\min_{q\in\Phi} KL(p||q)$. If $p \in \Phi$, then the projection reduces to an identity mapping. However, in general, $p \notin \Phi$ and hence such a projection is nontrivial.

Similar to MPA, EPA can be implemented on the factor graph in an iterative message passing manner. There are two steps for each round of iteration: FN update and VN update, respectively. For the VN update at the $t$-th iteration, the message $I^t_{k\to l}(\mathbf{x}_k)$ from VN $\mathbf{x}_k$ to FN $f_l$ is computed, and for the FN update, the message $I^t_{l\to k}(\mathbf{x}_k)$ of the opposite direction is computed. According to the principle of EPA, the messages are updated as follows

$$I^t_{k\to l}(\mathbf{x}_k) = \frac{\mathrm{Proj}_{\Phi}\left(p^t(\mathbf{x}_k)\right)}{I^{t-1}_{l\to k}(\mathbf{x}_k)}, \tag{6}$$

$$I^t_{l\to k}(\mathbf{x}_k) = \frac{\mathrm{Proj}_{\Phi}\left(q^t_l(\mathbf{x}_k)\right)}{I^t_{k\to l}(\mathbf{x}_k)}, \tag{7}$$

where

$$p^t(\mathbf{x}_k) = I_{\Delta\to k}(\mathbf{x}_k) \prod_{n\in V(k)} I^{t-1}_{n\to k}(\mathbf{x}_k), \tag{8}$$

$$q^t_l(\mathbf{x}_k) = I^t_{k\to l}(\mathbf{x}_k) \sum_{\mathbf{x}_m, m\in F(l), m\neq k} p(\bar{\mathbf{y}}_l|\bar{\mathbf{x}}_l) \prod_{m\in F(l), m\neq k} I^t_{m\to l}(\mathbf{x}_m), \tag{9}$$

and $I_{\Delta\to k}(\mathbf{x}_k) = P_0(\mathbf{x}_k)$ is the prior probability of $\mathbf{x}_k$ in (3).

If the projection set $\Phi$ is chosen to be Gaussian distribution, then the messages $I^t_{k\to l}(\mathbf{x}_k)$ and $I^t_{l\to k}(\mathbf{x}_k)$ reduce to Gaussian distributions which can be fully characterized by the mean value and covariance matrix. As a result, the computational complexity is significantly reduced. Moreover, as shown in (5), since the $l$-th likelihood FN is only related to $x_{kl}, k \in F(l)$, the messages of $I^t_{k\to l}(\mathbf{x}_k)$ and $I^t_{l\to k}(\mathbf{x}_k)$ can be further simplified to scalar complex Gaussian distributions, i.e., $I^t_{k\to l}(x_{kl})$, and $I^t_{l\to k}(x_{kl})$, respectively.

Specifically, at the $t$-th iteration, if the message $I^{t-1}_{l\to k}(\mathbf{x}_k)$, i.e., $I^{t-1}_{l\to k}(x_{kl})$, from the likelihood FN $f_l$ to VN $\mathbf{x}_k$ is circularly complex Gaussian with mean $\mu^{t-1}_{l\to k}$ and variance $\xi^{t-1}_{l\to k}$, i.e.,

$$I^{t-1}_{l\to k}(x_{kl}) \propto CN(x_{kl}; \mu^{t-1}_{l\to k}, \xi^{t-1}_{l\to k}). \tag{10}$$

From (8), we can compute $p^t(\mathbf{x}_k)$ as

$$p^t(\mathbf{x}_k) \propto I_{\Delta\to k}(\mathbf{x}_k) \prod_{n\in V(k)} CN(x_{kn}; \mu^{t-1}_{n\to k}, \xi^{t-1}_{n\to k}). \tag{11}$$

Then, to compute the message from VN $k$ to FN $f_l$, the $p^t(\mathbf{x}_k)$ is projected to a Gaussian distribution $CN(x_{kl}; \mu^t_{kl}, \xi^t_{kl})$ by matching the mean $\mu^t_{kl}$ and the variance $\xi^t_{kl}$ with respect to (w.r.t.) $p^t(\mathbf{x}_k)$. That is, the mean $\mu^t_{kl}$ and the variance $\xi^t_{kl}$ are computed w.r.t the approximated posterior probability $p^t(\mathbf{x}_k)$. Note that to compute the mean $\mu^t_{kl}$ and variance $\xi^t_{kl}$, normalization of $p^t(\mathbf{x}_k)$ is needed to obtain the discrete probability for $\mathbf{x}_k$ from the hybrid discrete and continuous density function in(11). In specific, suppose that the modulation size is $M$ and the codebook is $\boldsymbol{\mathcal{X}}_k$ of size $M$, then

$$p^t(\mathbf{x}_k = \boldsymbol{\alpha}) = \frac{P_0(\mathbf{x}_k)\prod_{n\in V(k)} CN(x_{kn} = \alpha_n; \mu^{t-1}_{n\to k}, \xi^{t-1}_{n\to k})}{\sum_{\boldsymbol{a}\in\mathcal{X}_k} P_0(\mathbf{x}_k)\prod_{n\in V(k)} CN(x_{kn} = \alpha_n; \mu^{t-1}_{n\to k}, \xi^{t-1}_{n\to k})}, \tag{12}$$

where $\boldsymbol{\alpha} \in \boldsymbol{\mathcal{X}}_k$ is an $L$-dimensional vector and $\alpha_n$ is the $n$th element of $L$-dimensional vector $\boldsymbol{\alpha} \in \boldsymbol{\mathcal{X}}_k$. In practice, the multiplications of Gaussian density functions in (12) can be computed in the log domain for simplicity. As a result, the mean $\mu^t_{kl}$ and the variance $\xi^t_{kl}$ are computed as

$$\mu^t_{kl} = \sum_{\boldsymbol{a}\in\mathcal{X}_k} p^t(\mathbf{x}_k = \boldsymbol{\alpha})\alpha_l, \tag{13}$$

$$\xi^t_{kl} = \sum_{\boldsymbol{a}\in\mathcal{X}_k} p^t(\mathbf{x}_k = \boldsymbol{\alpha})|\alpha_l - \mu^t_{kl}|^2. \tag{14}$$

From (7), we can compute the message $I^t_{l\to k}(\mathbf{x}_k)$ as

$$I^t_{l\to k}(\mathbf{x}_k) \equiv I^t_{k\to l}(x_{kl}) \propto CN(x_{kl}; \mu^t_{k\to l}, \xi^t_{k\to l}) \tag{15}$$

where

$$\frac{1}{\xi^t_{k\to l}} = \frac{1}{\xi^t_{kl}} - \frac{1}{\xi^{t-1}_{l\to k}}, \tag{16}$$

$$\frac{\mu^t_{k\to l}}{\xi^t_{k\to l}} = \frac{\mu^t_{kl}}{\xi^t_{kl}} - \frac{\mu^{t-1}_{l\to k}}{\xi^{t-1}_{l\to k}}. \tag{17}$$

Next, the message computation from FN to VN is described. Given the Gaussian messages $I^t_{l\to k}(\mathbf{x}_k)$ from VN to FN in (15), it can be derived that $q^t_l(\mathbf{x}_k)$ in (9) is also Gaussian which is denoted by

$$q^t_l(\mathbf{x}_k) \equiv q^t_l(x_{kl}) = CN\left(x_{kl}; \hat{\mu}^t_{kl}, \hat{\xi}^t_{kl}\right). \quad (18)$$

In fact, the mean and variance $\hat{\mu}^t_{kl}, \hat{\xi}^t_{kl}$ are precisely the linear MMSE estimates of $x_{k\square}$ over the linear equation(4), where the message $I^t_{l\to k}(\mathbf{x}_k)$ can be viewed as the prior Gaussian distribution of $x_{kl}, k \in F(l)$. Specifically, at the $l$-th RE, the prior mean and covariance matrix of $\bar{\mathbf{x}}_l = [x_{kl} | k \in F(l)]$ can be denoted as $\mathbf{m}^{\text{pri}}_l = [\mu^t_{k\to l} | k \in F(l)] \in \mathbb{C}^{d_f \times 1}$ and $\mathbf{\Sigma}^{\text{pri}}_l = \text{diag}([\xi^t_{k\to l} | k \in F(l)]) \in \mathbb{C}^{d_f \times d_f}$ , respectively, then the posterior mean $\mathbf{m}^{\text{post}}_l = [\hat{\mu}^t_{kl} | k \in F(l)] \in \mathbb{C}^{d_f \times 1}$ and covariance $\mathbf{\Sigma}^{\text{post}}_l \in \mathbb{C}^{d_f \times d_f}$ of $\bar{\mathbf{x}}_l$ can be calculated as

$$\mathbf{m}^{\text{post}}_l = \mathbf{m}^{\text{pri}}_l + \mathbf{\Sigma}^{\text{pri}}_l \bar{\mathbf{H}}_l^H \left(\bar{\mathbf{H}}_l \mathbf{\Sigma}^{\text{pri}}_l \bar{\mathbf{H}}_l^H + \sigma^2 \mathbf{I}\right)^{-1} \left(\bar{\mathbf{y}}_l - \bar{\mathbf{H}}_l \mathbf{m}^{\text{pri}}_l\right), \quad (19)$$

$$\mathbf{\Sigma}^{\text{post}}_l = \mathbf{\Sigma}^{\text{pri}} - \mathbf{\Sigma}^{\text{pri}}_l \bar{\mathbf{H}}_l^H \left(\bar{\mathbf{H}}_l \mathbf{\Sigma}^{\text{pri}}_l \bar{\mathbf{H}}_l^H + \sigma^2 \mathbf{I}\right)^{-1} \bar{\mathbf{H}}_l \left(\mathbf{\Sigma}^{\text{pri}}_l\right)^H. \quad (20)$$

The diagonal elements of $\mathbf{\Sigma}^{\text{post}}_l$ constitute the posterior variances of $x_{kl}$: $\left[\hat{\xi}^t_{kl} | k \in F(l)\right]$. Given the posterior mean $\hat{\mu}^t_{kl}$ and variance $\hat{\xi}^t_{kl}$ of $x_{kl}$, from (7) and (18), the messages $I^t_{l\to k}(\mathbf{x}_k)$ are also Gaussian, i.e., $I^t_{l\to k}(\mathbf{x}_k) \equiv I^t_{l\to k}(x_{kl}) = CN(x_{kl}; \mu^t_{l\to k}, \xi^t_{l\to k})$, whose mean and variance are [19]

$$\frac{1}{\xi^t_{l\to k}} = \frac{1}{\hat{\xi}^t_{kl}} - \frac{1}{\xi^t_{k\to l}}, \quad (21)$$

$$\frac{\mu^t_{l\to k}}{\xi^t_{l\to k}} = \frac{\hat{\mu}^t_{kl}}{\hat{\xi}^t_{kl}} - \frac{\mu^t_{k\to l}}{\xi^t_{k\to l}}. \quad (22)$$

Then, the approximated posterior probability $p^t(\mathbf{x}_k)$ in (12) can then be updated as $p^{t+1}(\mathbf{x}_k)$ using the $\mu^t_{l\to k}, \xi^t_{l\to k}$ in (21)-(22). Till now, the $t$-th iteration is completed. After multiple iterations, the posterior LLRs for coded bits $c_{k,j}$ can be calculated based on the updated $p^{t+1}(\mathbf{x}_k)$. After subtracting the prior LLRs, we obtain the extrinsic LLRs feedback to the FEC

$$\lambda^{\text{e}}\left(c_{k,j}\right) = \log \frac{\sum_{\alpha \in \mathcal{X}^+_{k,j}} p^{t+1}(\mathbf{x}_k = \boldsymbol{\alpha})}{\sum_{\alpha \in \mathcal{X}^-_{k,j}} p^{t+1}(\mathbf{x}_k = \boldsymbol{\alpha})} - \lambda^0\left(c_{k,j}\right), \quad (23)$$

where $\boldsymbol{\mathcal{X}}^+_{k,j}$, $\boldsymbol{\mathcal{X}}^-_{k,j}$ denote the set of NOMA constellations of UE $k$ with $c_{k,j} = 1$ and $c_{k,j} = 0$, respectively.

Summarizing the above discussion, the proposed universal EPA detector is shown in *Algorithm* 1, where $T_{\text{inner}}$ is the number of maximum inner-loop iterations of EPA. Note that only the MU detector in Fig. 1 is presented in *Algorithm* 1.

From *Algorithm 1*, it can be seen that the main complexity of EPA detector lies in the matrix inversion whose order is $O(N_r^3 L)$. The complexity is linear with respect to the spreading factor $L$ since EPA is implemented in a chip-by-chip manner, i.e., the MMSE equalization operation can be performed on each RE independently.

In the progress of NOMA standardization, EPA has already been adopted as candidate receivers for many NOMA schemes with sparse RE mapping, e.g., SCMA [20] and PDMA [21], as a simplified version of MPA receiver. However, as *Algorithm 1* shows, the implementation of EPA does not rely on the sparsity in the transmitted signals and is universal for all potential NOMA schemes. However, as will be shown in [16], having sparsity design in resource mapping can help to reduce the inter-user interference in a NOMA system and thus improve convergence and reduce implementation complexity.

***Algorithm1: EPA Detector***

·**Initialization**:

(a) Calculate the prior distribution $P_0(\mathbf{x}_k)$ via the feedback LLRs from the FEC decoder. Initialize $p^1(\mathbf{x}_k) = P_0(\mathbf{x}_k)$, for $k = 1, \dots, K$.

(b) Initialize the mean and variance from FN $f_l$ to VN $\mathbf{x}_k$ as $\mu^0_{l\to k} = 0, \xi^0_{l\to k} = \infty, l = 1, \dots, L, k \in F(l)$.

·**Iteration: For $t = 1$: $\mathbf{T_{inner}}$, do**

**VN Update**: For $k = 1, \dots, K, l \in V(k)$

1) Compute the posterior mean $\mu^t_{kl}$ and variance $\xi^t_{kl}$ as (13) and (14).
2) Update the extrinsic mean $\mu^t_{k\to l}$ and variance $\xi^t_{k\to l}$ from VN to FN as (16) and (17).

**FN Update**: For $l = 1, \dots, L$

1) Perform chip-by-chip MMSE as (19) and (20)
2) For $k \in F(l)$, update the mean $\mu^t_{l\to k}$ and variance $\xi^t_{l\to k}$ from FN to VN as (21)and (22)

**Posterior Probability Update**:

For $k = 1, \dots, K$, update $p^{t+1}(\boldsymbol{x}_k = \boldsymbol{\alpha})$ via (12).

·**LLR Calculation:**

Compute the extrinsic LLRs via (23).

## III. Experimental Results

In this section, link level simulations of the proposed EPA receiver for both spreading based and non-spreading based NOMA schemes are presented and discussed. The spreading based NOMA scheme is exampled by frequency domain spreading (FDS) with the NOMA signatures listed in Table A-2 of [22] with spreading length $L = 4$. The non-spreading based scheme, on the other hand, is exampled by contention based OFDMA (CB-OFDMA), which there is no change in the transmitter side design compared with the traditional OFDMA, but the receiver is changed to the turbo-like iterative receiver to accommodate multi-user transmitting on the same resource.

The traditional MMSE detector with codeword level PIC is also evaluated for comparison (MMSE receiver in short). Note that all the receivers are implemented in a chip-by-chip manner so that the complexity will not scale in cubic relation with the spreading length.

The channel model follows the TDL-A model [23] with delay spread 30ns and the moving speed is 3km/h. For FEC encoding, LDPC code is adopted. The simulation results are averaged over 5,000 transmissions. In all simulations, the number of outer-loop iterations for the turbo-like structure in Fig.1 is $T_{\text{outer}} = 3$, and the number of inner-loop iterations for EPA receiver is $T_{\text{inner}} = 3$.

Fig. 3 shows the BLER performances of FDS versus signal to noise ratio (SNR) for $K = 6, 8$ users with $N_r = 2$ receiver antennas at payload size equals to 60, 40 bytes. It can be seen from Fig. 3 that the traditional MMSE-PIC receiver (MMSE receiver in short) could not reach 10% BLER for {6 UEs, 60 Bytes} case under the settings. On the contrary, the EPA receiver can greatly improve the convergence such that 6 UEs with 60 bytes payload size can be supported. It also can be seen that the EPA receiver can help FDS to improve the BLER performance by 2.5 dB for {8 UEs, 40 Bytes} case.

Fig. 4 shows the BLER performances of CB-OFDMA versus SNR for $K = 6, 8$ users with $N_r = 2$ receive antennas and 75 bytes and 60 bytes payload size, respectively. The EPA receiver can help improve the convergence and thus help enhance the BLER performance by 1.5 dB and 5.2 dB for {6 UE, 75 Bytes} and {8 UE, 60 Bytes} cases, respectively.

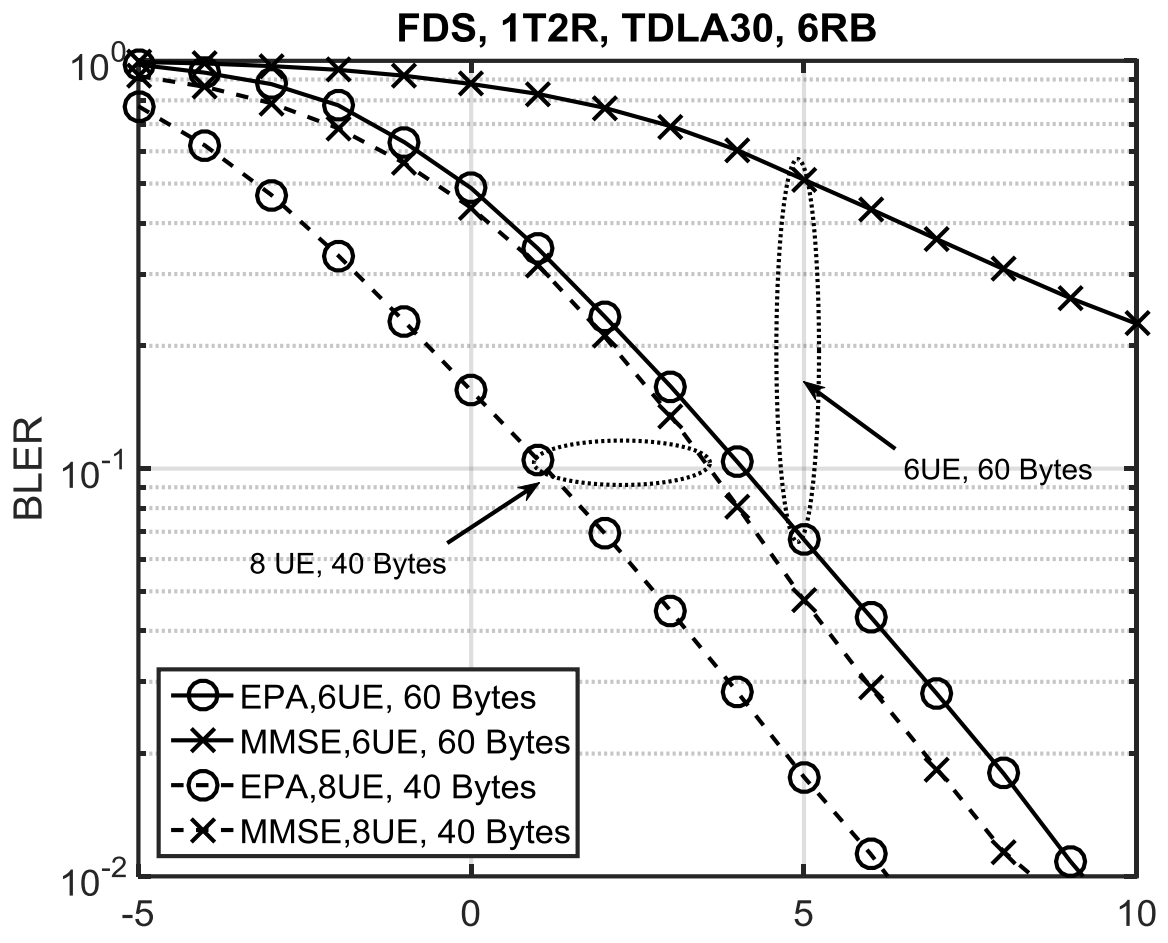

Fig. 3. Performances comparison of frequency domain spreading (FDS).

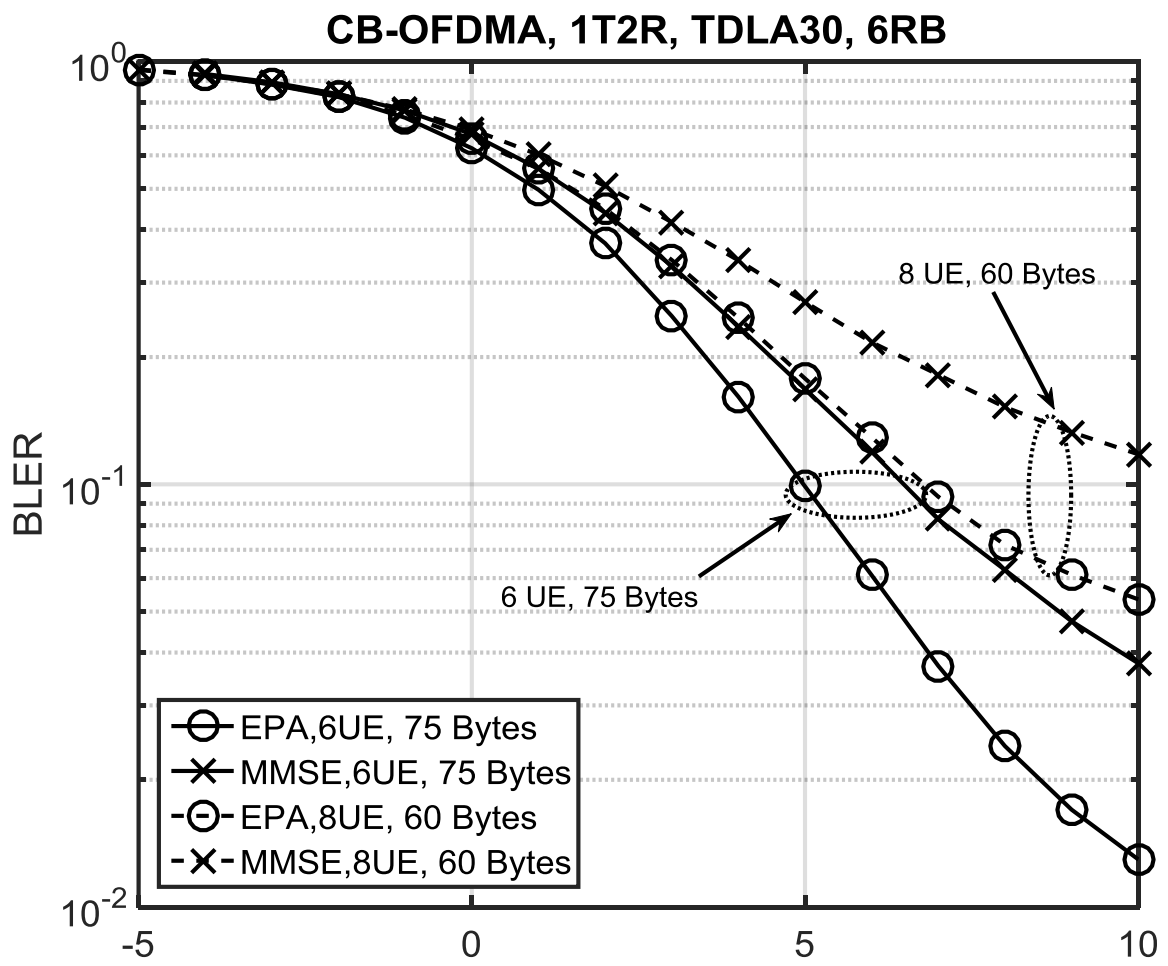

Fig. 4. Performances comparison of CB-OFDMA.

## IV. CONCLUSION

In this paper, a universal EPA receiver is proposed under a general turbo-like iterative receiver structure, which is suitable for all potential uplink NOMA schemes. The proposed EPA receiver has better BLER performance than the traditional codeword level MMSE-IC receiver with only linear complexity w.r.t. modulation order, number of users multiplexed on the same resource, and the spreading factor, respectively.